# Implementing an insect brain computational circuit using III-V nanowire components in a single shared waveguide optical network.

David O. Winge[1*], Steven Limpert[1], Heiner Linke[1], Magnus T. Borgström[1], Barbara Webb[2], Stanley Heinze[3], Anders Mikkelsen[1*]

[1]Department of Physics & NanoLund, Lund University, P.O. Box 118, 221 00 Lund, Sweden

[2]School of Informatics, University of Edinburgh, 10 Crichton Street, Edinburgh EH8 9AB, UK

[3]Lund Vision Group, Department of Biology, Lund University, 22362 Lund, Sweden

*email: david.winge@sljus.lu.se, anders.mikkelsen@sljus.lu.se

**Recent developments in photonics include efficient nanoscale optoelectronic components and novel methods for sub-wavelength light manipulation. Here, we explore the potential offered by such devices as a substrate for neuromorphic computing. We propose an artificial neural network in which the weighted connectivity between nodes is achieved by emitting and receiving overlapping light signals inside a shared quasi 2D waveguide. This decreases the circuit footprint by at least an order of magnitude compared to existing optical solutions. The reception, evaluation and emission of the optical signals are performed by a neuron-like node constructed from known, highly efficient III-V nanowire optoelectronics. This minimizes power consumption of the network. To demonstrate the concept, we build a computational model based on an anatomically correct, functioning model of the central-complex navigation circuit of the insect brain. We simulate in detail the optical and electronic parts required to reproduce the connectivity of the central part of this network, using experimentally derived parameters. The results are used as input in the full model and we demonstrate that the functionality is preserved. Our approach points to a general method for drastically reducing the footprint and improving power efficiency of optoelectronic neural networks, leveraging the superior speed and energy efficiency of light as a carrier of information.**

# Introduction

The neural computation performed by real brains remains an important inspiration for machine intelligence. However, software implementations of artificial neural networks using standard computer hardware are orders of magnitude less energy efficient compared to biological brains[1,2], limiting future applications. To address this challenge, a multitude of physical/chemical mechanisms such as memristors[3], ionic liquids[4] and spintronics[5] are being explored to realize naturalistic neural networks[6]. Recently, the use of photonics based solutions has gained renewed interest[7,8] as it can overcome both speed and efficiency limits of standard technology for neural networks[7,9–12]. For bio-inspired processing networks, a main energy expenditure and complexity challenge is in the need for a large number of communication connections between components[9,13]. Using light for network connectivity is in principle a superior solution as it can transmit information quickly and with high energy efficiency. However, realizing the full potential of optical solutions is hindered by their large circuit footprint and the energy losses in regular (macroscopic) optoelectronic components.

Significant progress has been made in concentrating and manipulating light using nanostructure components, thus allowing for the necessary miniaturization of optical computation circuitry. In particular, III-V nanowires have matured into a versatile, controllable and well characterized nanotechnology platform. This has allowed the development of novel light harvesting[14–16] and emission technologies[17,18] as well as combination with Si based technology. III-V heterostructure nanowires can uniquely be tailored with widely varying optical and electronic properties. They respond locally and efficiently to optical signals, concentrate light on a sub-wavelength scale[19,20], and have a natural polarization sensitivity that has been used for optical logical gates[21]. Importantly, they can have a much higher absorption cross-section than their physical size[20,22] and can thus act as efficient photodetectors. Precise and varied large scale 2D arrays of functionalized nanowires[20] and single nanowire emitters with controllable emission patterns [17,18,23,24], as summarized in ref. 25, have been manufactured and experimentally studied in detail.

An excellent way to explore the potential of III-V nanostructured components for neural networks is to implement specific circuit models based on a detailed understanding of biological neural circuits. The insect brain offers substantial advantages as a target, as its lower complexity and higher accessibility supports functional understanding at the single neuron level. At the same time, insects are capable of tasks well beyond the reach of current artificial neural nets, such as traveling across hundreds of kilometers of unfamiliar terrain to pinpoint a specific breeding ground[26,27], or returning to a near invisible nest entrance from several kilometers away in a straight-line trajectory, after a convoluted searching trip through dense vegetation[28]. Using only a few drops of nectar as energy supply, they achieve all this with a brain the size of a grain of rice, which contains ca. 100000 times fewer neurons than mammalian brains.

One module of the insect brain conserved across species with vastly different lifestyles is the *central complex* (CX), which is a core decision making and motor control circuit[29,30]. The neural circuit of the CX has been decoded in great detail, which is of the utmost importance to any attempt to mimic the neural functionality. It is characterized by tight structure-function coupling, in which the anatomical layout of a circuit defines its computations. One important purpose of this neuronal circuit is to serve as a navigational control system that underlies most planned, directed movements of insects[29]. The CX has been distilled to its fundamental neuroarchitecture and the function of a number of its components was mapped onto a biologically constrained computational model[31]. This model has the ability to integrate the outward going path of a simulated insect leaving its nest and to switch into producing the required steering signals to enable the insect to navigate directly back to its point of origin. This homing task is successfully carried out using input of limited precision and with considerable circuit noise. Containing less than 100 neurons of qualitatively similar function, the central complex model is simple enough to serve as a target system for investigating novel nanotechnology solutions for neural networks, while still being important for solving real navigational tasks in insects.

In this paper, we describe how the spatial and energy footprint of an optical neural network that reproduces a key part of the insect central-complex circuit can be minimized using nano-components placed inside a shared waveguide. We first describe the model of the central complex that we implement and the general requirements of the nodes and their network inter-communication architecture. These principles should be widely applicable in reproducing any neural circuit. Second, a nanowire-based device is shown to be a prime candidate for the neural node as it can have a very small energy consumption and large cross-section for light detection. Third, optical simulations on the network level demonstrate how the inter-device coupling weights are set by emission patterns and geometrical layout, inside a shared quasi 2D waveguide. This broadcasting strategy is a key component of our design, as it reduces the spatial footprint of the network, removing the need for both inter-node connecting waveguides and inter-device electrical wiring. Fourth, the results from the electrical and optical modelling is tested by substituting it into the full computational model of the insect brain central complex circuit[31], successfully demonstrating that the navigation capability is preserved. Finally, we evaluate the operational efficiencies needed in order to realize our optoelectronic implementation.

# Results

### General concept of the neural network and its implementation

To establish the basic design criteria for our hardware solutions, we provide a brief discussion of the insect brain neural navigation network model of Stone et al[31] which is the foundation for exploring and demonstrating our approach. Converted into mathematical form and implemented on a standard computer, the CX network allows an insect to be guided back to its nest after a foraging trip ("the insect" in this case is an abstract agent in the computer that receives input data from either an artificial or a real environment; the model has been shown to work for a real world robot[31]). The model use the insect's current heading and speed as input, and by integration generates an internal (vector) representation of the angle and distance of the point of origin. Once homing is initiated, the same circuit outputs a left or right steering signal that indicates how the insect should change its heading in order to move homeward. The model can perform this task with limited, noisy input data and deals successfully with obstacles blocking its path. It can function with internal noise levels in the neural processing of up to 20%.

In the lower right corner of Fig. 1 we show the three main network layers of this navigational circuit, which can be represented topologically as concentric circles. The innermost layer and the heart of the CX model is a ring attractor circuit, which constantly keeps track of the heading of the insect. This layer receives its input from specialized compass neurons, as schematically indicated top right of Fig. 1. Each ring attractor neural node communicates both inwards with is peers, as well as *outwards* with the subsequent layers of the circuit. This adds a recurrent feature to this layer. The second layer is a memory layer, which receives input from both the ring attractor and speed input neurons (not drawn in Fig. 1). Using this information, it performs a path integration to keep track of the home direction. The third layer compares the current heading direction from the ring attractor with the desired heading towards home, given by the memory neurons, to compute the steering signal. This description summarizes features of the network relevant for the present paper, for further details and biological justification of the circuit design refer to ref. 31.

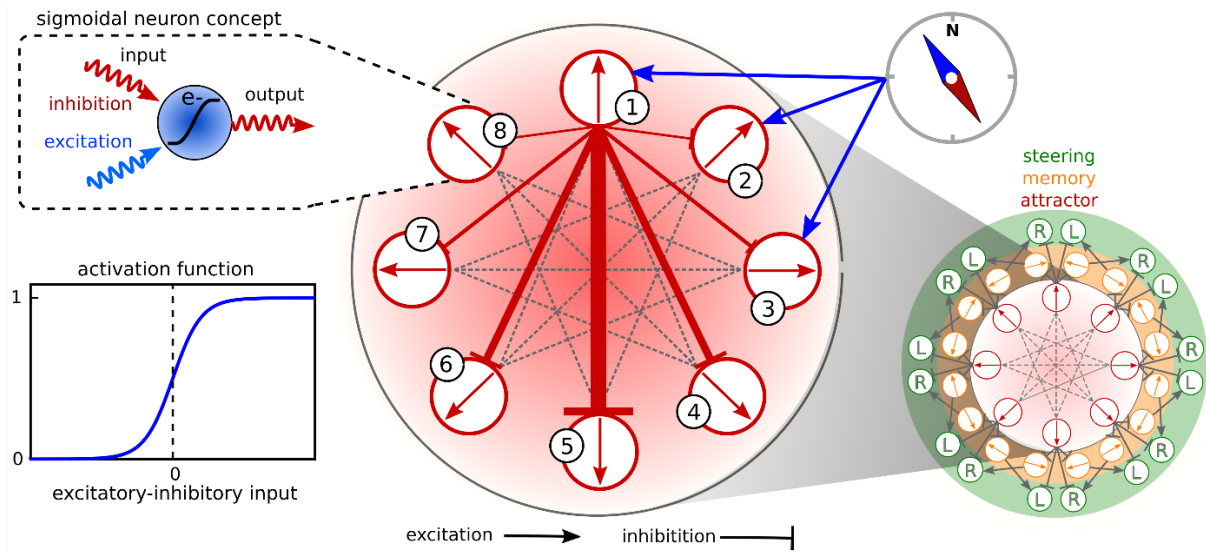

**Fig 1. The ring attractor network, that is implemented in this study, which is the most connected sub-circuit of the insect brain central complex model (CX) of ref. 31.** The CX neural network main parts (schematically shown in bottom right) can be represented in a circular topology as three concentric and interconnected ring network layers (attractor, memory and steering). The central ring attractor network layer is enlarged and shown in the center of Fig. 1. It is the focus of the present study. Each of the eight neural nodes (neurons) in the ring attractor is shown as a circle with an arrow, in turn representing the direction of the insect. All ring attractor neurons are mutually inhibitory and at the same time provide information to the outer memory and steering ring network layers. For clarity we only show the inhibitory connections from node 1 to the other nodes in detail (red lines ending in a bar) in the central part of Fig 1. The weights are exemplified by the thickness of the red lines, being strongest for the directly opposing node. The existence of the other inter-connecting channels are indicated as dashed gray lines. The input to each node is given by external compass neurons as exemplified by the blue arrows. In the top left corner we show the schematic activation function of the sigmoid neural node. It receives inhibiting and exciting signals and output either an inhibiting or exciting signal of similar kind. Inset in bottom left corner shows an example the sigmoidal electronic data processing of each neural node. All nodes in the CX have this functional behavior, however the offset and slope varies.

In the present study, we focus on the innermost ring attractor layer which has the largest and most complex connectivity pattern. The nodes in this sub-circuit perform signal evaluation in a qualitatively similar fashion to all other nodes across the network. Implementing this inner ring will therefore demonstrate and test the main aspects of our approach. This requires the design of an optoelectronic component that can serve as a node as well as designing a network of these nodes that fulfils the interconnection weight requirements.

The artificial neuron (neural node component) to be constructed is a sigmoidal neuron entity as schematically shown in the top left inset of Fig. 1. Sigmoidal neurons operate using a rate code, i.e., the frequency of neuronal action potentials is encoded as a continuous numerical value. This generic neuron type will receive external inhibitory and excitatory input from multiple sources. It may also have a constant internal input source (for example a bias), which creates an offset in the activation function. All inputs are weighted, added and the sum evaluated via a non-linear sigmoidal function, which will result in activation of an output signal (if the excitation sufficiently dominates inhibition) that must then be transferred to several other neurons. For biologically inspired neural implementations of the node, it is important that both the slope and offset of the activation function can be tuned. In our hardware implementation the neural firing rates of both input and output signals are represented by light intensities (rate of photons) while the sigmoid evaluation is performed electrically within the neural node component.

A topological illustration of the full ring attractor is provided in the center of Fig. 1. Each of the eight neurons in the ring attractor network are represented by a circle with an arrow that in turn represents a specific directional heading of the insect. In the specific CX architecture, excitatory inputs provide compass and speed information, while the neurons communicate via inhibitory signals among themselves. The strongest inhibition is from the neural node on the opposite side of the ring, gradually falling off for neighboring nodes. This stabilizes the activity in the inner ring to a single bump centered on one of the neurons[31]. This is how the insect obtains a robust sense of direction.

To achieve this weighted communication, the neural node components will be distributed within a single planar waveguide in a circular pattern (resembling the topological layout of the circuit, see Fig. 1) and the light emission patterns of the nodes will be shaped using the morphology of the nano-optoelectronic/photonic structures. As the communication between the neural nodes is achieved inside a single shared waveguide, the inhibiting and excitatory signals must operate alongside at different wavelengths. The waveguide defines the plane of computation: each node responds to all available local optical signals and emits the appropriate nonlinear response into the waveguide anew. This approach allows us to use the propagation direction normal to the waveguide plane for supplying input signals as well as for probing of the network.

### Realization of the sigmoidal neural node component

In this section we propose a specific III-V nanowire component to implement the sigmoidal neuron and present simulations demonstrating that it meets the design criteria outlined above. In brief, this component should receive (inhibiting/exciting) optical signals, weigh them, process the result through a suitable (sigmoid) activation function and send out an optical output (inhibiting/exciting) signal. As each node receives optical signals broadcast through the shared waveguide channel, it naturally absorbs many different signals from many sources. The multitude of collected signals represent a clear analogy to the dendrites of biological neurons.

We propose a T-shaped component, as shown in Fig. 2, to perform these tasks. To be practically feasible, the design is as simple as possible, and is based on existing III-V nanowire technologies. It consists of a main nanowire with two npn photo-transistors (for input) and a branch with a nanowire LED (for output). The two input npn transistors are engineered in two material systems; InP and $Al_{0.3}In_{0.7}P$ to achieve different wavelength responses for exciting and inhibiting signals. Using a realistic physical model, with parameters from experimental studies, we show that the individual parts of the component can work under realistic conditions. By mapping these results onto circuit elements as shown in Fig. 2b, the complete component is modeled and it is shown to provide adequate activation functions for neural processing as will now be described.

The two colored regions (blue/red) in Fig. 2a indicate material regions absorbing at different wavelengths, where the excitatory signal ($\lambda_+$) is carried by a shorter wavelength than the inhibitory signals ($\lambda_-$), so $\lambda_+ < \lambda_-$. Upon receiving a signal of $\lambda_+$, the excitatory photo-transistor gives rise to a net current through the LED (when no inhibition is present). If there is an inhibitory signal at wavelength $\lambda_-$ simultaneously present, the current generated by the inhibitory photo-transistor is subtracted from the excitatory current. The branched circuit thus sums the two currents, where the inhibitory signal is weighted with a negative sign. Furthermore, the resistance in series with the LED ensures that the load line is practically flat with respect to bias, leading to a saturation above a certain current $I_{\text{sat}}$. The ideal mathematical operation in this case is

$$I_{\text{LED}} = min(I_{\text{sat}}, I_+ - I_-), \quad I_+ > I_- \geq 0.$$

which constitutes a basic nonlinear activation function. The equivalent circuit diagram of the component is shown in Fig. 2b summing inhibitory and excitatory signals and evaluating them via the (sigmoid) function to provide the output. This will be used to simulate the exact behavior of the component, which will deviate from the ideal case.

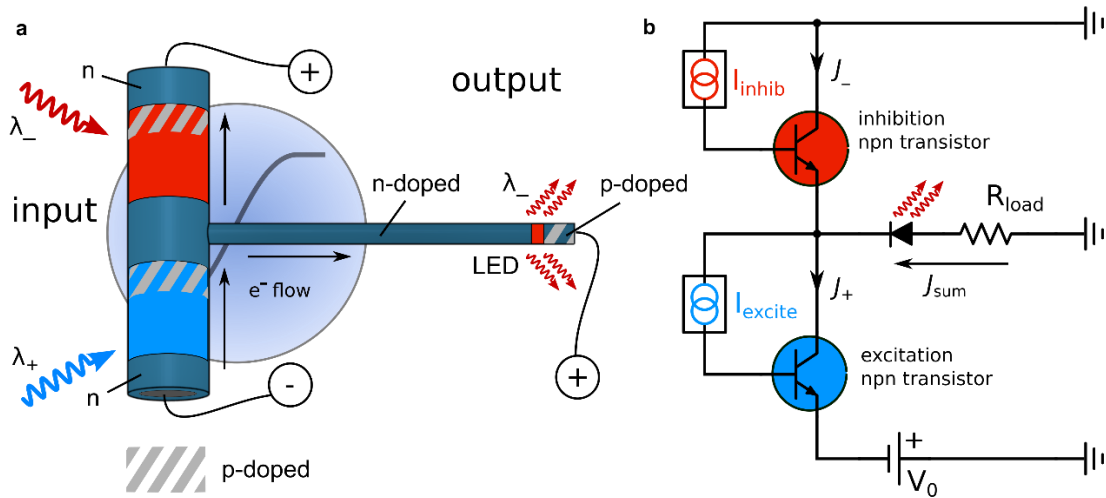

**Fig. 2. Diagrams of the nanowire based artificial neuron that will evaluate input light signals, resulting in an appropriate light output**. a) Schematic of the neural node component which is a branched nanowire with two npn transistors in the stem and a LED in the branch. The red and blue parts indicate III-V materials of small/large bandgaps respectively. p-doped regions on the nanowire is indicated by the grey stripped regions, the remaining areas are of the nanowire are n-doped. Electrical contacts needed to power the device are indicated. The length and diameter for the thick main nanowire are 700 nm and 200 nm, respectively, while for the branch the corresponding dimensions are 1000 nm and 50 nm. b) Equivalent circuit model of the device in a) with the floating base npn photo-transistors modeled in a common collector configuration with current sources representing the generated excitatory and inhibitory photo-currents.

Our simulation is based on specific III-V nanowire implementations for high efficiency photovoltaics[16,20,33]. It thus assumes that the two phototransistors are fabricated through heterostructure nanowire growth and selective doping along the principal axis. A low bandgap material, as indicated with red color in Fig. 2a, effectively traps the holes of the photo-current, while the electrons are separated out by the collector *pn*-junction, defined by selective n and p-doping. The LED branch can be grown in the same material system as the phototransistors, with an undoped emission section defined by a thin layer of InP, centering a *pn*-junction defined by selective doping, again along the principal axis. The material region absorbing at $\lambda_-$ is susceptible also to the excitatory $\lambda_+$ signal - to avoid significant signal contamination experimentally verified wavelength and polarization specific nano-antennas[32] can be used to focus the external excitatory light input on the $\lambda_+$ region. For a detailed description of the phototransistor, LED design and modeling we refer to the Materials and methods section and supplementary information (SI).

The main results from modeling the circuit of Fig. 2 are presented in Fig. 3. Fig. 3a shows a band diagram of the npn phototransistor $\lambda_-$ wavelength receiver in the floating base configuration. Here it is possible to distinguish the low band gap region of the base and collector in between the two graded heterojunctions at about 100 and 300 nm, respectively. The wide gap emitter effectively captures the holes generated by the photo-current, while the electrons are swept away by the base-collector *pn*-junction. In Fig 3b, the current through the LED under excitation and inhibition is shown, exemplifying the results from the circuit modelling of the complete component. The inset displays two examples of activation functions used by different kinds of neurons of the full network model[31]. They are characterized by their slope and offset (inversion point) which varies for different types of neurons. Our component can reproduce these varying functional features. The slope depends both on the bias over the circuit as well as on design parameters such as base region length, emitter-base bandgap offset and doping[34] and the saturation current is directly controlled by the load resistance (see the SI for examples). As seen in the inset of Fig. 3b, our device naturally produces a result similar to the activation function of the memory layer neurons of ref. 31. A zero point offset need to be added to replicate the behaviour of the ring attractor neurons which can be accomplished by either adding a constant background input signal or a bias unit as described in the SI.

In Fig 3b we show that the activation function is stable under a significant range of inhibiting currents and close to a sigmoid function in terms of shape. The effect of a changing slope for higher inhibitory currents seen in Fig 3b is a consequence of the non-ideal elements of the circuit. When simulating the complete network function (see below), no degradation of the performance due to this effect can be found.

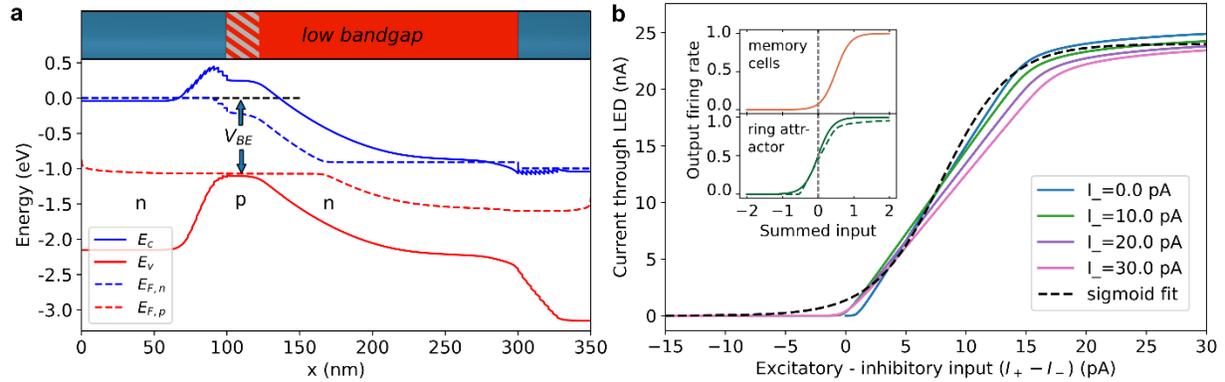

**Fig. 3. Results from the electronic modeling of the III-V neural node component.** a) Band diagram of one of the two npn phototransistors at an applied bias of 1 V. Above the diagram the corresponding regions of the npn transistor from Fig 2a are indicated. Solid lines represent band edges and dashed lines quasi-Fermi levels. Here the wide-gap emitter is depicted on the left side and the collector to the right. b) Results from modelling of the full circuit in Fig. 2b) with the parameters $V_0 = 3.0$ and $R_{\text{load}} = 30$ MΩ. Current through the LED as a function of the difference in excitatory and inhibitory current for different fixed values of inhibitory currents is shown. A fit to a sigmoid function is added for comparison. Inset shows two different activation functions from Stone et al.[31]. Dashed line in ring attractor inset shows the renormalized nanowire component activation function for comparison.

### Realization of communication between neural node components in the ring attractor

We now demonstrate how the neural node components can communicate via optical signals through a shared waveguide structure, and how the signals are weighted to produce the necessary pattern of coupling coefficients (as discussed above and indicated in Fig. 1). In this pattern, the component directly opposite from an emitting node should receive the maximum signal intensity, which should then fall off gradually towards the closest components. Any self-inhibition of the node component should be minimized. To achieve these conditions a dipole emitter is suitable. The dipole source serves the additional purpose of transmitting signals in the 2D plane outwards from the center. This facilitates the necessary communication of the ring attractor nodes with the nodes in the two outer ring network layers (indicated in Fig. 1 bottom right).

As shown in Fig. 4a,b (to scale), the components are positioned in a geometry directly inspired by the topographic representation of the ring attractor central network shown in the center of Fig. 1. The inhibition nanowire branch of each neural node face inwards, and the nodes are placed inside a HfO$_2$ "guiding" layer (as depicted in Fig. 4b) which keeps the light in the 2D plane of the components. The SiO$_2$ substrate / HfO$_2$ / air structure thus represents a quasi 2D waveguide. The weights are determined by the emission pattern of the LED emitter of each neural node component. InP nanowires have been reported to have giant polarization anisotropy [35]. As a result, the coupling of light from inside of the nanowire to its surroundings depend on the polarization. The large dielectric contrast between the nanowire and its surrounding material strongly favors coupling to light fields polarized parallel to the nanowire axis. The dipole emission pattern corresponding to this polarization is shown in Fig. 4c. For InP nanowires surrounded by air, the polarization ratio was calculated as $\rho = 0.96$ (ref. 35), so nearly all intensity is spread in a pattern as indicated in Fig. 5b, where the whole device geometry is taken into account in the calculated emission. In the SI we discuss the dipolar emission from InP nanowires embedded in the waveguide structure in more detail.

To simulate the absorption and emission between the components and calculate the corresponding coupling weight matrix $g_{ij}$, we constructed a full 3D model of the optical network with its node components in the commercially available FDTD solver from Lumerical[36]. FDTD methods are widely used to model nanowire optical absorption[22], scattering[37], emission[18,38], demonstrating good agreement with experimental observations.

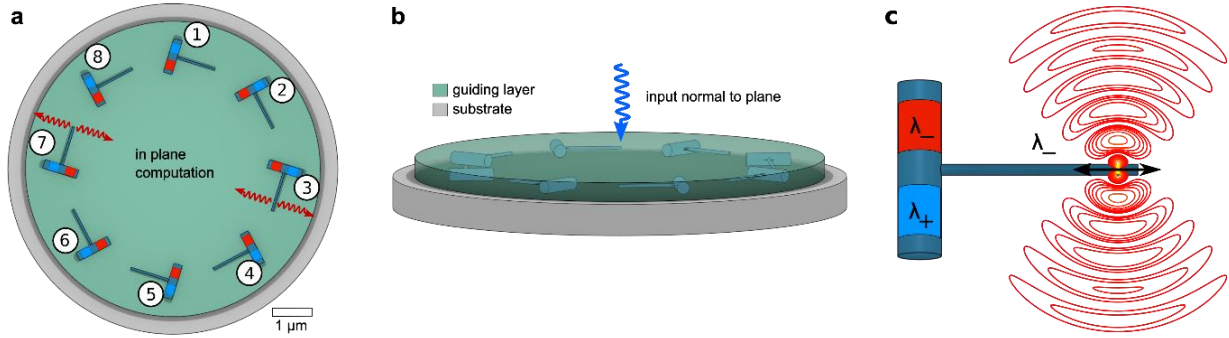

**Fig. 4. Drawing to scale of the inner ring attractor network using the neural node component described in the previous subsection.** Panel a) and b) shows a top and side view of the network, encapsulated in a waveguide structured as $SiO_2/HfO_2$/air. Scale bar is given in a. The drawings are radially cropped. In the full network design the waveguide would continue outwards, creating a semi-infinite waveguide in the radial direction. The internal inhibitory signaling occur in the 2D plane of the network system as seen in a. The external excitatory light (compass) input in the $\lambda_+$ region is to enter perpendicular to the 2D communication plane of the network structure as seen in b. To ensure that these signals reach the correct input position wavelength and polarization specific nano-antennas[32] can be used to focus. In c) one device is shown together with an emission pattern of a dipole source oriented along the nanowire branch. The two absorption regions are again indicated by color as in Fig 2.

We determine the absorption of each device by calculating the optical transmission through a closed box around each absorption region in a component. The fraction of intensity absorbed in device (i) relative to the emitted power of device (j) directly corresponds to the weight matrix indices $g_{ij}$ displayed in Fig. 5c. The intensity flowing out of the waveguide in both the horizontal and lateral directions was recorded to calculate the waveguide confinement factor $\Gamma$, found to be around 60% at 830 nm. The remaining 40% is light lost from the communication processes. Stand-alone modelling containing one single device was also performed to understand how the thick receiver branch interferes with dipole emission. The resulting intensity emission pattern w(θ) is shown in Fig. 5b demonstrating that the main dipole shape of the radiation is retained.

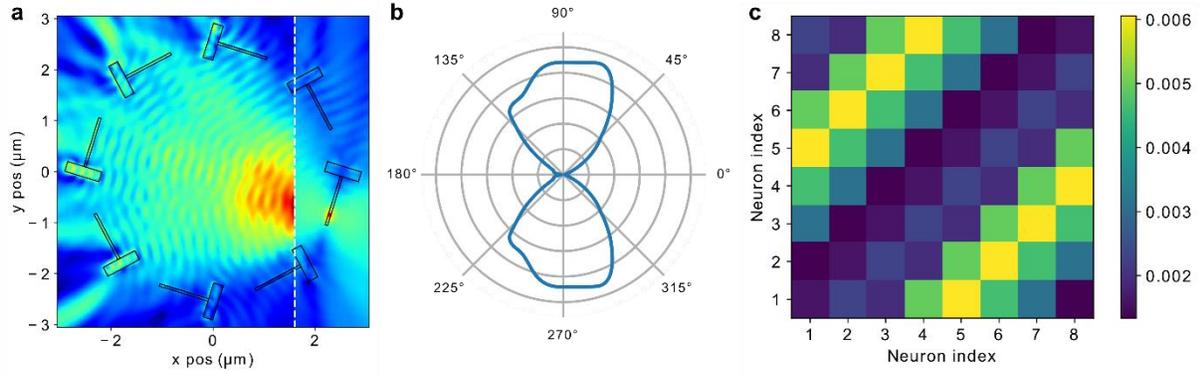

**Fig. 5.** a) The field distribution of dipole emission from device 3 (as denoted in Fig 4a) in the xy-plane of the waveguide based on FDTD modelling. To the right of the dashed white line, a log scale is used to visualize the fields close to the dipole source. b) Polar emission pattern from the radiating dipole inside the nanowire branch (oriented horizontally in the figure). c) Inter-device weight matrix calculated from the absorption in each device, with labels as in Fig. 1 and 4. An asymmetry can be seen due to the rotation of the components as seen in Fig. 4a.

The absorption region of a device is subject to some radiation from its own emitter (as seen in Fig 5a), so that $g_{ii} \neq 0$. It is important to minimize any such re-absorption because it leads to undesirable *self-inhibition* in the component. For the network to perform well, it is enough that $g_{ii}$ is substantially smaller than the inter-device coupling coefficients $g_{ij}$. This can be achieved using the polarization selectivity of the dipole source and extending the LED branch to move the light emitter away from the receiver branch. Increasing the length of the emitter branch however increases the footprint and asymmetry of the device, but a sweet spot can be found that optimizes all these parameters sufficiently for the circuit to function well. These considerations set the length of the emitter branch used in the device design throughout the paper.

From the top view the asymmetry of the component in the circular pattern leads to an overall asymmetry in the network weights. For example, the distances from the emitter of device node 1 (see Fig. 4a) to device nodes 4 and 6 are not identical. This is reflected in the matrix plot of the modeled interconnecting weights in Fig. 5d). We show that this has no significant influence on the navigational capacity of the network in the following section.

## Simulation of the full navigation network using the III-V nanowire based ring attractor

We have parametrized the neural node component and the optical network in terms of activation functions and inter-connecting weights. Using the complete computational model of the insect navigation CX[31], we can now test the performance of the network when using III-V nanowire components. In detail, we replace the activation functions in the ring attractor (inset of Fig. 3b) with the results obtained from the circuit simulation (full plot in Fig. 3b). In addition, the communication weights connecting the ring attractor components were replaced by the simulated counterparts of the matrix seen in Fig 5c. For use in the computational model, the output current (for input current see Materials & Methods), as well as the weight matrix values, were normalized to unity. The slope-saturation discussed in connection to Fig. 3b was fully taken into account, by making the activation function explicitly depend on both exciting and inhibiting currents and not only their sum.

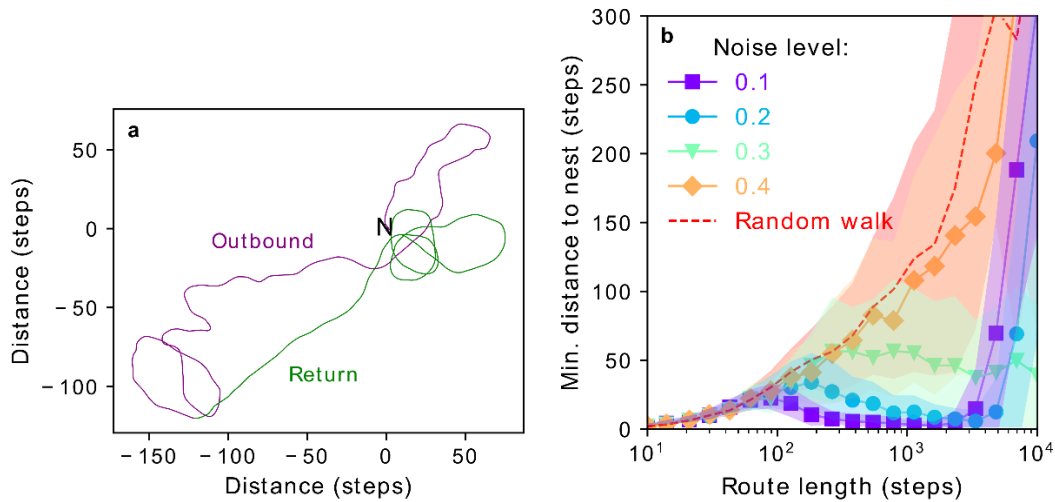

**Fig. 6. Simulated navigation using the model described in Stone et al.[31] combined with the specific III-V system proposed here.** a) An example route using the simulated III-V nanowire device results integrated into in the full computational model. First the insect performs a random foraging trip (purple line). At a given point it is switched to return home to the nest indicated by N (green line). When it reaches the nest it will keep circling it as can also be seen in the green line trace. b) Statistics showing the success rate for 1000 traveled routes distributed on 20 different trip lengths for each noise level. The noise is added to the result processed by each activation function (final value clipped to the interval [0,1]), both inside and outside the ring attractor. This number corresponds to the amplitude of the white noise that was added to

the signal which was in turn normalized to unity. The standard deviation is depicted as filled areas around the respective lines. A reference case using a completely random walk for homing is also shown for comparison.

The results of the navigation tests are summarized in Fig. 6 using the physical parameters from the III-V components placed in the network. In Fig. 6a an example route with 1500 steps is shown where the agent finds its way back without difficulty. This is the case with added noise of 0.1. In Fig. 6b the results of a statistical study where the signal noise was systematically increased up to 0.4 is shown. In summary, the network is capable of handling trips of a maximum of 5000 steps and a noise level of 0.2 before the agent starts to have troubles with finding its way back. These results are on par with the ones presented in Stone et al.[31] and represents a clear success of the network with the new III-V components.

## Operational efficiencies of the III-V optoelectronic network implementation

In each step of transmission, detection and signal processing in the network, energy is dissipated, either due to conversion losses or intensity leaking out of the waveguide. To counteract the power dissipation and achieve stable operation, enough built-in signal amplification is required in the output of each neural node to drive the subsequent input in other nodes (fulfilling demands on fan-out and cascadability[39]). We discuss the different efficiencies in the process, evaluate their magnitudes based on our calculations and experimental values to demonstrate the feasibility of our approach and estimate necessary efficiency limits.

Starting from the optical input signals in the III-V neural node component, the photon-exciton conversion efficiency (antenna efficiency $\eta_a$) describes how efficiently the photons absorbed in the device generate electron-hole pairs in the base and collector regions of the two npn photo-transistors. The efficiency of converting the electronic signal back to photons in the LED branch (the internal quantum efficiency (IQE), $\eta_{IQE}$) describes the relative effectiveness of the radiative recombination compared to competing processes such as trap-assisted and Auger recombination. Then the light extraction efficiency $\eta_{out}$ factors with the IQE to provide the external quantum efficiency of the emitter. To make up for the lost power in these processes and balancing the network so that each neural node component provides the necessary output to achieve fan-out of input to other receiving nodes, a built-in current amplification factor $\beta$ of the output is needed.

The output power of device (i) in terms of absorbed power, process efficiencies and amplification then reads:

$$P_{\text{out,i}} = \beta\, \eta_{\text{out}} \eta_{\text{IQE}} \eta_{\text{a}} P_{\text{abs,i}}.$$

The absorbed power ($P_{\text{abs,i}}$) consists of contributions from all other node components as well as exterior (compass) input. For a given component (i), the power contribution from component (j) can be calculated using the geometrical coupling coefficient $g_{ij}(r)$, describing the overlap of the emission light pattern of node component (j) and the absorption cross-section of receiving component (i). For cascadability and fan-out to be fulfilled, each node component must be able to deliver at least enough of $P_{\text{abs}}$ to each of the other components in order to activate them. Demanding that component (j) must be able supply the full power needed for component (i) constitute an upper limit of power needed, so we can set:

$$g_{ij} P_{\text{out},j} \geq P_{\text{abs,i}},$$

Finally noting that all neural node components are identical (the output powers of (i) and (j) will be similar in size) we can combine all of this to the inequality:

$$g_{ij} \eta_{\text{out}} \eta_{\text{IQE}} \eta_{\text{a}} \geq \frac{1}{\beta}$$

where we relate the losses and efficiencies to the current gain of the npn bipolar photo-transistors. Most of the values in this equation can be determined from our modelling and experimentally known values. For the ring attractor, the strongest geometrical coupling coefficient is $g_{ij} \approx 0.006$ (as shown in Fig. 5c). This efficiency includes the waveguide loss and describes how much of the total signal that reaches a connected component with the largest weighting factor. The internal quantum efficiency is closely related to the trap-assisted recombination lifetime of 1.34 ns used here for the III-V nanowires. From the device modelling (further detailed in the SI) we find a maximum efficiency of $\eta_{\text{IQE}} = 0.7$ for a current density of $J = 800\ \text{A/cm}^2$. This number is well beyond the low-current limit for npn bipolar photo-transistors where $\beta$ saturates[40]. Regarding the antenna efficiency, it has been shown recently that InP nanowires with a diameter of 310 nm, designed for solar cells, enjoy a photo-carrier collection efficiency of 90% over several microns[41].

The diameter is similar to the one in the sketch shown in Fig. 4 which motivates our choice of $\eta_a = 0.9$ for this estimate. As a final step the current gain factor is set to $\beta = 1900$ based on the component modeling results. This leads us to a final estimate for the required light extraction efficiency from the LED

$$\eta_{\text{out}} > \frac{1}{g_{ji}\, \eta_{\text{IQE}}\, \eta_a} \frac{1}{\beta} \approx 0.14.$$

This is well below the number of 42% as reported in ref. 18. So our model appears feasible given realistic values from modelling and experiments. However, as discussed below, further work will be important for improving the energy efficiency of the system.

## Discussion

We have proposed and successfully simulated an optoelectronic design to implement the nodes and connections of a neural circuit, closely based on the insect brain, that carries out an important navigational task. A relevant question is if all components are available for a practical realization of the proposed III-V nanostructure implementation. III-V nanowire bipolar photo-transistors have been developed and evaluated for photovoltaics and detectors[42]. Nanowire based emitters that can perform above the required efficiencies have been experimentally realized[17,18]. Individual nanowires have been demonstrated to have a light concentration factor of 8 (ref. 22) and high efficiency nanowire solar cells[20] have been realized. The nanowire branch which constitutes the LED can be grown using bottom-up techniques as described in previous studies[43], or realized by crossed nanowires[44].

A network that requires many device nodes with similar operating parameters is a challenge to realize in most nanotechnologies. However, the present analog computational device has a large robustness built into the architecture. This is shown in the statistical outcome of the navigation tests that was performed on the network and shown in Fig. 6. Here a signal noise of 20% can be tolerated before the results got significantly worse than the noise-free reference. In addition, the effect of inaccuracy of the positioning of each component is estimated in the Materials and Methods section, addressing the deviations in rotational precision. We show that deviations of up to 13° can be tolerated for the largest coupling coefficient. This indicates that significant variations in device positioning and perfection can be tolerated. It is relevant to note that a wide variety of microscopy based diagnostic tools are available today for optimization of the optical fields and electron excitation locally in III-V nanowire structures[45,46].

We do not view the emitter polarization engineering as a major obstacle in constructing the components. The InP nanowires have a giant polarization anisotropy[35], which naturally helps shape the optimal emission pattern shown in Fig. 5b. Embedding the nanowire components in a waveguide decreases the dielectric contrast and reduces the anisotropy. To re-enhance this anisotropy, one option is to use a tapered nanowire cavity[18] which has the additional benefit that it, through the Purcell effect, decreases the spontaneous emission lifetime in the quantum dot. This directly leads to a better efficiency of the emitter $\eta_{IQE}$ which translates to lower operating currents and power consumption. Better emission control also allows a decrease in the network diameter and thus increase the geometrical coupling coefficients $g_{ij}$. Among other possible solutions[47], the antenna structure demonstrated by Ramezani et al.[48] is suitable for controlling the emission from the nanowire branch emitter.

The shared waveguide design allows us to skip inter-component wiring or waveguides, and instead set the weights using the geometry of the system. This strategy is very beneficial to achieve a small footprint and low energy use. Further generalizing this concept to different networks might require additional design developments and new light focusing components. However, a wide variety of sub-wavelength nanophotonic structures have been designed recently that can guide light to focus in specified points with varying intensity. Wiring to supply power to the active components is still required and these will cause some additional scattering. ITO can be used to minimize this scattering and we do not foresee this as a major obstacle.

Before finally discussing the power consumption of our network solution, we would like to put it in context by briefly relating it to biological systems and CMOS technology. The human brain is known to operate at 10-20 W and based on simple assumptions the energy consumption per neuron and operation has been estimated at $10^{-16}$ J (see ref 1). More detailed studies of the energy consumption of the neural system in the brain have been put forward, however, estimates end up in a similar range[49]. Exactly how the brain spends this energy is a matter of debate, but it has been estimated that around 70% is used for inter-neuron communication (ref. 50). Using CMOS solutions particularly optimized towards neural networks, efficiencies in the at $10^{-11}$ J per operation range have been achieved[2,51]. This is already considerably better than standard computers, but orders of magnitude below the brain.

The power dissipation bottleneck for the present design is the nanowire LED efficiency. In order for the total losses not to overcome the transistor gain factor, the emitter must be operated at a relatively high external quantum efficiency. As an example, a moderate internal efficiency of 50% requires a current through the LED of about 100 A/cm$^2$. This corresponds to ~2 nA in the branch. Assuming that a few volts is applied across the circuit and additional energy dissipation due to possible inhibition, a reasonable estimate is ~10 nW per neural node component during operation. The energy needed per operation depends on the frequency, but with experimentally verified values for components and reasonable assumptions on operation, an energy dissipation of $10^{-16}$ J/operation or less can be reached (see SI for more details), equivalent to the levels observed in biological brains. To further reduce the power consumption, the most important optimization is the LED efficiency at lower currents. If the trap-recombination lifetimes could be increased towards bulk values of InP, we expect an improvement of one or two orders of magnitude.

In conclusion, we have investigated two major new concepts for an artificial neural network system based on nanoscale optoelectronics. First, optical communication is done directly via broadcasting with all components in the same 2D slab confining the radiation. This radically reduces the footprint since no wiring (electrical or fiber) between components is needed. Second, we use a mature III-V nanowire technology platform to create the neural nodes. The nanowires have light absorption cross-sections much larger than their geometric dimensions and the III-V materials are very efficient in photon-electron conversion. To investigate their feasibility, we implemented these concepts on the most heavily interconnected part of a specific, anatomically verified model of the navigation center of the insect brain. This allowed a thorough simulation of all electrical and optical parts of the network using experimentally verified parameters. Using conservative estimates for all parameters and already available nanowire technology we show that the network will function and can be orders of magnitude more efficient compared to present technologies.

While the present work can be viewed as a proof of principle, it also identifies challenges for the development of such networks in terms of device design. Central is the power efficiency of the artificial neuron. The more efficient emission and absorption of light in the nanowire components, the more favorable solutions become. Another important challenge is the placement of the components and the focusing of light in sub-wavelength structures. For placement, technology relevant for other applications such as III-V nanowire based quantum computers face similar challenges and have put forward several solutions. The focusing and manipulation of light on a sub-wavelength scale has seen a wealth of new developments in recent years, thus creating even advanced patterns that can act as low footprint communication paths is possible. Again, energy dissipation is an important issue as many of such components have significant losses.

The ring attractor system that we implemented is in principle dedicated to a specific navigational task. Importantly, its functional connectivity can be expressed in geometrical terms that allow us to exploit light broadcast as a method of internode communication. The extent to which this may be a general principle in biological neural networks is unknown, but the ring attractor itself appears to subserve a wide range of navigational functions for the insect. As such, the methodology we have described might have greatest application for reproducing specific, but crucial, capabilities of biological brains. On the other hand, our proposed nano-scale nonlinear processing unit with optical input and output may serve as the minimal unit in many other neural network approaches.

## Materials & Methods

### Electronic modeling of the sigmoidal component

For a detailed account of the modeling of the devices we refer to the SI, but provide a short summary here. We use a drift-diffusion model with thermionic emission boundary conditions implemented in COMSOL to calculate the transport in the nanowire devices. This have been shown to yield good agreement to experimental data for InAs nanowire heterostructures[52] and InP *pn*-junctions[53]. The devices studied here are heavily doped and the main effects due to the surface is the increased carrier recombination due to surface states, why no explicit surface charge was considered here. We use an effective 1D model where the surface recombination term enters as an additional trap-assisted recombination process[54]. The modeling is divided into two steps. In the first, we model the nanowire npn photo-transistor and the LED. We fit the results to an Ebers-Moll model and the Shockley diode equation, respectively. In the second step, we use them as elements in the equivalent circuit of our device as shown in Fig. 2b). Here the two current sources $I_{\text{inhib}}$ and

$I_{\text{excite}}$ represent the photo-induced current in the base and collector regions of the respective transistors, operated in a common-collector mode. A large resistance $R_{\text{load}}$ ensures that the load line is essentially flat with respect to the bias. Using a spice solver the final results are extracted as shown in Fig. 3b).

This trapping of holes increases the optical gain of the npn phototransistors[34]. However, the fast trap and surface recombination in nanowires strongly limits the gain and the functionality. In this study we use a realistic electron and hole recombination lifetime of 1.34 ns as measured at room temperature in ref. 55, together with the experimentally observed mobilities in nanowires listed therein. Despite these limitations, we show that nanowire npn photo-transistors can deliver current gain factors $\beta > 1900$ which is needed to transmit the signals across the ring attractor.

For the nanowire LED the fast recombination process prevents a high efficiency at low currents, as a large density is needed for the spontaneous emission process to compete to non-radiative processes. In this work the momentum matrix element for the spontaneous emission process is calculated from the Kane energy of InP of 20.7 eV[56].

In order to use the simulated activation function in the computational model, the output current through the LED was normalized by the saturation current $I_{\text{sat}}$, while the input currents were normalized instead by $I_{\text{sat}}/\beta$ to take the amplification in each npn phototransistor into account. This directly yields the activation function shown as a dashed line in the ring attractor inset of Fig. 3b.

## Optical modeling of the network

In the FDTD 3D model, the ring attractor network of 8 devices was placed inside a guiding layer of 300 nm $HfO_2$, surrounded by $SiO_2$ and air, as shown in Fig. 4a,b. This quasi-2D waveguide confines 60% of the intensity emitted by the devices inside the network, which would otherwise suffer dramatic losses. It has been designed for a wavelength of 830 nm, which matches the bandgap of InP which we use for the absorbing region (red) and for the recombination region of the nanowire LED. A dipole emitter, representing the nanowire LED, was placed 100 nm from the end of the thin nanowire in device 3, oriented along the axis of the nanowire branch. The network circumference, with respect to the center of gravity of the wide nanowire of each device, was set to

$$2\pi R_{network} = 2L_{branch}N_{devices}$$

in order to leave generous space between components for wiring. This resulted in a network of size $2R_{network} = 5.1$ μm, with a branch length of $L_{branch} = 1.0$ μm. The InP and SiO$_2$ was modeled using the data of ref. 57, the HfO$_2$ using ref. 58, while the wide-gap material Al$_{0.3}$In$_{0.7}$P was modeled as a dielectric with refractive index n(AlInP) = 3.3, as its dispersive properties are of little interest in this study.

Each device was rotated 0.3 rad in the clockwise direction around the waveguide normal, as seen in Fig. 4a. Rather than having all main nanowires pointing to the center, this was done in order to enhance the opposite coupling coefficient (i.e. $g_{15}$) and to reduce the coupling to the clockwise neighbor (i.e. $g_{12}$). It is possible to estimate the effect of positioning noise relative to this rotation. Focusing on the strongest weights $g_{15}$, the ideal dipolar radiation pattern $cos^2\theta$ can be expanded around its peak at $\theta = 0$. This yields a direct relation of the uncertainty in angle corresponding to a certain relative uncertainty in the radiation pattern, namely

$$(\Delta\theta)^2 = p,$$

where $p$ now denotes the relative uncertainty. For example, it has been shown that the CX can navigate successfully with 5% weight noise. Using $p = 0.05$, we find a corresponding uncertainty in angle of $\Delta\theta = 13°$, which indicates that some positioning noise in the positioning of the components is tolerable. Expansions far from the peak are less favorable, giving a smaller allowed uncertainty for coupling coefficients of devices closer to each other. However, the absolute value of these coefficients is smaller in general which limits the impact of positioning errors in these couplings.

## Data availability

The data that support the plots within this paper and Supplementary Information are available from the corresponding author upon request.

## Acknowledgments

This work was supported by the Swedish Research Council (VR) and NanoLund FutureThemes.

## Conflicts of interest

The authors declare no competing financial interest.

## Contributions

AM conceived the project and initiated the work. AM and DOW designed the component and the network with important input from SL, HL, MTB, BW and SH. DOW performed the simulations, analyzed the data, optimized the component and network, and prepared the figures. AM and DOW wrote the manuscript with significant contributions from all other authors. All authors engaged in discussions and analysis of the results.

# Supplementary Information to: Implementing an insect brain computational circuit using III-V nanowire components in a single shared waveguide optical network

David O. Winge, Steven Limpert, Heiner Linke, Magnus T. Borgström, Barbara Webb, Stanley Heinze, Anders Mikkelsen

## I. MATERIAL MODEL

Here we detail the material model used for the drift diffusion modeling made in COMSOL. The two semiconductor materials InP and $Al_{0.3}In_{0.7}P$ were defined using the parameters in Table S1. The band properties are determined at 300 K from the recommended values in [1]. As AlInP is direct up to 0.44 Al fraction, we use the AlP Γ-point for the interpolation of band gap and valence band offset. To find an average heavy hole mass with respect to the different crystal directions we apply the spherical approximation [2], by replacing the Luttinger parameters $\gamma_3$ and $\gamma_2$ with their average value. The conductivity and density of states (DOS) effective hole masses are then calculated by the standard procedure of weighting their contributions to the conductivity and effective densities, respectively. As a realistic estimate we assume that the lifetime for the minority carriers are the same in the mixed material system as compared to InP. The mobility can then be estimated using the effective masses as

$$\mu = \frac{2\pi e}{\gamma m_{e/h}^{\text{cond}}}, \qquad (1)$$

where $\gamma$ is the electron scattering rate reported in [3], assumed to be the same in both materials.

| Property | InP | $Al_{0.3}In_{0.7}P$ |
|---|---|---|
| Band gap $E_g$ | 1.35 eV | 2.11 eV |
| Valence band offset | -0.94 eV | -1.18 eV |
| Dielectric constant (static) | 12.4 | 11.62 |
| Electron eff. mass $m_e$ | $0.08m_0$ | $0.12m_0$ |
| Hole DOS eff. mass $m_h^{DOS}$ | $0.75m_0$ | $0.76m_0$ |
| Hole cond. eff. mass $m_h^{cond}$ | $0.55m_0$ | $0.55m_0$ |
| Electron mobility $\mu_n$ | 490 cm$^2$/Vs | 320 cm$^2$/Vs |
| Hole mobility $\mu_p$ | 70 cm$^2$/Vs | 70 cm$^2$/Vs |
| Electron scattering rate $\gamma$ | 45 THz | 45 THz |
| Recombination lifetime $\tau_n$ | 1.34 ns | 1.34 ns |
| Refractive index (real) @ 830 nm $n$ | 3.45 | 3.3 |
| Refractive index (imag) @ 830 nm $k$ | 0.2 | 0.0 |

TABLE S1. Parameters used to define the two different materials in the COMSOL modeling. The bare electron mass is given as $m_0$.

We point out that no quantization effects were included when modeling the nanowire LED. Here we expect a slight blue-shift of the emission frequency due to the quantization shift. Such a shift would improve the sensitivity of the nanowire absorbers as the complex part of the refractive index is increasing with $\omega$, and at the same time increase the self-inhibition effect.

For the FDTD simulations the refractive index data for InP was given by [4] while we model the $Al_{0.3}In_{0.7}P$ segments as a simple dielectric with $n = 3.3$ fixed at the value at 830 nm, as our region of interest is far from the bandgap of $Al_{0.3}In_{0.7}P$ at 2.11 eV (590 nm). The optical characteristics of the mixed material system have been measured [5] and the reported results suggest that interpolation is reasonable in this spectral region. The value of $n$ for $Al_{0.3}In_{0.7}P$ is based on linear interpolation between InP and AlP data from [4] and [6, 7], respectively.

## II.   MODELING OF THE NPN PHOTOTRANSISTORS

The npn phototransistors for inhibition and excitation are designed following [8]. Both feature low bandgap base and collector regions, with a p-doped base and n-doped emitter and collector, according to the values in Tab. S2, resulting in a band diagram as shown in as shown in Fig. S1. Here, for the inhibition npn phototransistor, the low bandgap material is InP and the high bandgap material $Al_{0.3}In_{0.7}P$. For the excitation npn phototransistor, the low bandgap material needs to have a slightly higher bandgap relative to the inhibition phototransistor, in order to be transparent to the output from the InP nanowire LED. For example, the InP region could be replaced with $Al_{0.1}In_{0.9}P$, which shifts the bandgap with respect to wavelength from 918 nm to 767 nm. This should be sufficient for a simulated emission peak width of less than 100 nm for the InP nanowire LED, as shown in Fig. S4b). For this feasibility demonstration, we assume that this small difference in the low bandgap material leads to small differences in the fitted parameters and use the same transistor model for the inhibition and excitation npn phototransistors. This helps clarify the influence of each component in the circuit modeling.

| Property | Value |
|---|---|
| Temperature | 300 K |
| Base doping | $n_A = 5 \cdot 10^{18}/\text{cm}^3$ |
| Emitter doping | $n_D = 2.5 \cdot 10^{18}/\text{cm}^3$ |
| Collector doping | $n_D = 1.0 \cdot 10^{17}/\text{cm}^3$ |
| Matrix element $E_P$ (InP) | 20.7 eV |
| LED doping | $n_{D/A} = 1.0 \cdot 10^{18}/\text{cm}^3$ |
| Doping junction depth | 25 nm |
| $\tau_{\text{Auger}}$ | $1.0 \cdot 10^{-30} \text{cm}^6/\text{s}$ |

TABLE S2. Parameters used to define the COMSOL model together with the material data presented in Fig. S1.

The high bandgap material of the emitter section makes it transparent to the incoming radiation at the target wavelength, which increases the efficiency of the phototransistor as electron-hole pairs are generated predominantly where they contribute to the base current. The collector contact of our device is again engineered using the high bandgap material, as shown in Fig. 3a) of the main text. To prevent the formation of barriers between the base and the collector and emitter sections, respectively, a graded heterojunction of 30 nm length was used at both sides. The electron-hole pairs, generated in the base-collector region made up of the low bandgap material, are separated by the built in potential where the electrons are drained by the collector whereas the holes are stuck in the base and effectively lowers the gate potential of the base region. As described by [8], it is important to have a narrow base region to achieve a high current gain factor. To achieve this on the nanoscale we make sure that the p-doping in the base is higher than both the emitter and collector doping levels. This ensures a sharply defined base region which can then be limited to a narrow region, as shown in Fig. 3a) of the main text. For this design the base region was set to 20 nm and defined via a doping density of $n_A = 5 \cdot 10^{18}/\text{cm}^3$, compared to the emitter and collector contact regions having $n_D = 2.5 \cdot 10^{18}/\text{cm}^3$ and the collector $n_D = 1.0 \cdot 10^{17}/\text{cm}^3$. The main reason for the extra care taken in optimizing the base region is that the device is sensitive to the fast surface and trap-assisted recombination in the nanowires. This requires a strong current gain to compensate for the loss of carriers.

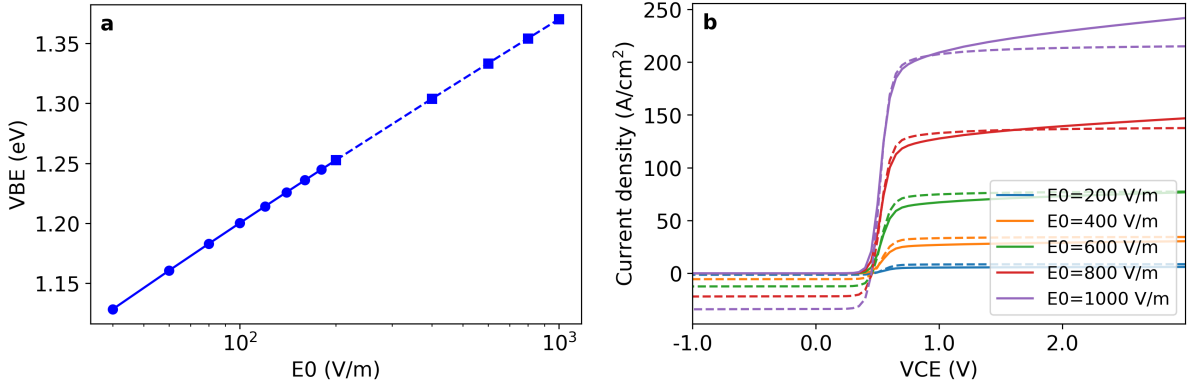

FIG. S1. a) The voltage drop $V_{\text{BE}}$, as depicted in Fig 3a) of the main text, extracted from the simulations at specific field strengths $E_0$, for a fixed $V_{CE} = 1$ V. b) Current densities in the base region of the inhibition npn phototransistor. The dashed lines display the hole current density, above the critical bias of 0.8 $V_{CE}$, multiplied with $\beta_F = 1900$.

A COMSOL 1D semiconductor model was constructed to perform a semi-classical simulation of the nanowire npn phototransistor. The results from this modeling were then mapped upon a refined Ebers-Moll transistor model. This allowed us to simulate the full neural node component, containing multiple nonlinear sub-components (npn phototransistors and an LED), through standard circuit simulation software. Here we describe first the transistor model of our choice, then present the details of the COMSOL modeling and finally demonstrate the fitting procedure that maps the modeling results onto a circuit element.

For a realistic circuit simulation of our nanowire based neuron, outside the COMSOL environment, a transistor model taking all relevant physical mechanisms into account is needed. An analytical model with few parameters that is still able to resolve all four regions of transistor operation is a refined Ebers-Moll model (similar to hybrid-pi models [12]), which

we write here as

$$J_C = J_S \left(e^{qV_{BE}/n_{be}kT} - e^{qV_{BC}/kT}\right)(1 + \gamma V_{CB}) - \frac{J_S}{\beta_R}\left(e^{qV_{BC}/kT} - 1\right) \quad (2)$$

$$J_B = \frac{J_S}{\beta_F}\left(e^{qV_{BE}/n_{be}kT} - 1\right) + \frac{J_S}{\beta_R}\left(e^{qV_{BC}/kT} - 1\right) \quad (3)$$

where $J_C$, $J_B$ are collector and base current densities, respectively, $\gamma = 1/V_A$, $V_A$ is the Early voltage and $n_{be}$ is the diode ideality factor for the base-emitter junction. The Early effect takes the base narrowing into account which is important for a short device such as ours, while the base-emitter ideality factor takes into account the significant recombination taking place in the base-emitter junction, due to the heterostructure design.

By formulating our Ebers-Moll transistor model with this few number of parameters (using for example the same saturation current density $J_S$ for all terms), we indirectly carry out a number of assumptions. We have neglected the influence of high-level injection at high $V_{BE}$ as well as the excess base current contribution to $J_B$. However, the model only requires fitting of the Early voltage $V_A$, the current gain in reverse $\beta_R$ and forward $\beta_F$ direction, the ideality factor $n_{be}$ and the saturation current density $J_S$. Few fit parameters provides a clear understanding of the underlying physics which is important in this type of feasibility study.

For the detailed modeling, a COMSOL 1D semiconductor model was constructed with the simulation parameters as listed in Tab. S2 and material parameters as listed in Tab. S1. For transport we use a semi-classical drift-diffusion model with thermionic boundary conditions [9, 10] and Fermi-Dirac statistics for the carriers. As the barriers of the photo-transistor have an effective thickness of about 50 nm, see Fig. 3a) of the main text, tunneling should not be significant [11]. The conduction and valence band are modeled in the effective mass approximation. For the trap-recombination, all trap states are considered to be at the same energy and optically generated carriers are assumed to follow the thermal distribution, i.e. they are instantly cooled. The spontaneous emission rate is calculated using the $E_P$ matrix element as

$$E_P = \frac{2m_0}{\hbar^2}\mathbf{P}^2, \quad (4)$$

where $\mathbf{P}$ is the quantity entering the COMSOL model. For details about the meshing sequence to ensure convergence, please contact the corresponding author.

The main results from the COMSOL simulations of the inhibition npn phototransistor are shown in Fig. S1. To be able to fit Eqs. (2)-(3) for the collector and base current it is important to supply $V_{BE}$ as illustrated in Fig. 3a) of the main text, for each value of the exciting optical field strength $E_0$. This can be done for a fixed collector bias $V_{CE}$ as shown in Fig. S1a) for $V_{CE} = 1$ V. Here $E_0$ is increased to generate a suitable span of optically induced $V_{BE}$. For a fixed field strength $E_0$, $V_{BE}$ saturates with increasing $V_{CE}$, above a critical voltage $V_c$. Below this value, $V_{BE}$ cannot be linked to a corresponding $E_0$ or vice versa. In Fig. S1a) this critical voltage can be estimated to $\sim 0.8$ V. In Fig. S1b) the current density is plotted close to threshold for the transistor. The base current is also plotted, multiplied with the forward gain factor that we find from the parameter fit discussed below.

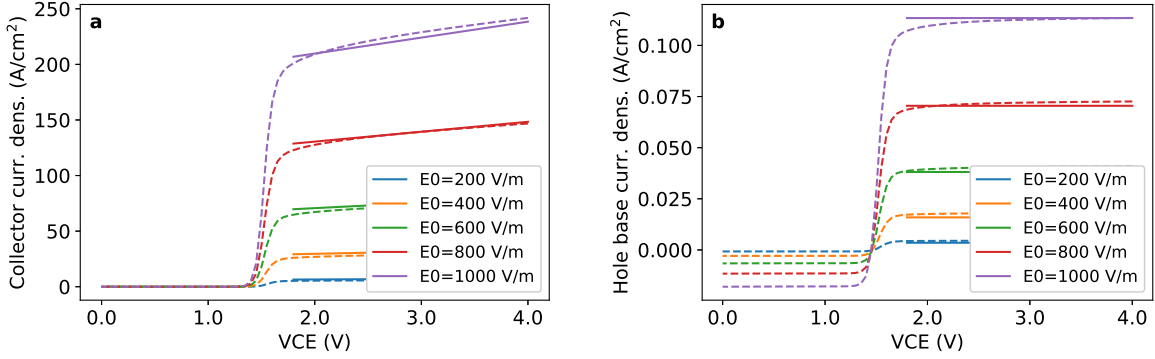

FIG. S2. a) Electron current density and b) hole current density fitted to a simple Ebers Moll model for constant $V_{BE}$ taken from our simulation results.

To find the Ebers-Moll parameters $J_S, \beta_F, \beta_R, n_{be}$ and $V_A$ we first find $V_A = 15.0$ V and $n_{be} = 1.3$ from inspection and fit the other parameters using a multi-variable fit using the Nelder-Mead simplex algorithm as implemented in the python `scipy` library. As the target function we use

$$R(J_S, \beta_F, \beta_R) = \sum_i \left( J_C^{\text{sim}}(V_{BE,i}) - J_C(V_{BE,i}, J_S, \beta_R) \right)^2$$
$$+ \beta_F^2 \left( J_B^{\text{sim}}(V_{BE,i}) - J_B(V_{BE,i}, J_S, \beta_F, \beta_R) \right)^2 \quad (5)$$

where $J_C$, $J_B$ are given by Eqs. (2)-(3). The forward gain factor is introduced to normalize the contributions from the hole and electron currents. The fit yielded the final parameters $\beta_F \approx 1900$, $\beta_R \approx 1.0$ and $J_S = 5.3 \cdot 10^{-16}$ A/cm$^2$.

With all the parameters in place, we are able to set up a benchmarking SPICE model to check how well it reproduces the phototransistor behavior. The transistor is modeled in the floating gate configuration by using a current source to represent the photo-generated base current in a common-collector configuration [13] as depicted in Fig. S3a), where we also add a load of 1 kΩ. As the excitatory current we use the base current density that we record from the COMSOL model and the results are shown in Fig. S3b). A good agreement is achieved in terms of onset voltage, magnitude and Early effect, which indicates that the physical model in Eqs. (2)-(3) captures the most important aspects of operation.

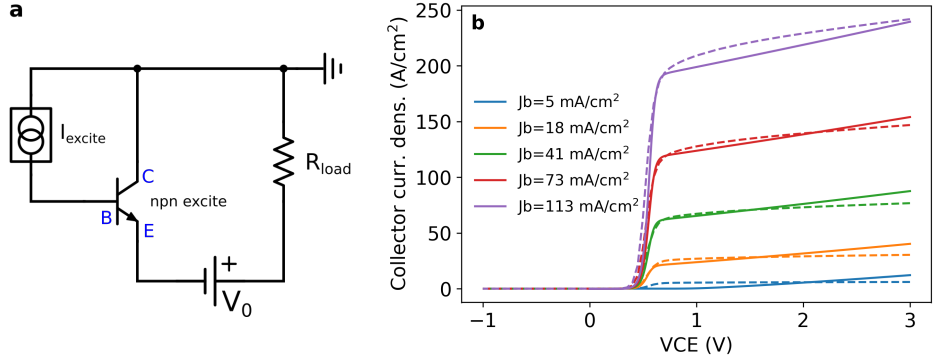

FIG. S3. a) Circuit model with a floating base configuration for the photo-transistor.. From the figure it is apparent that the transistor is operated with a common collector, with $V_0$ as the collector-emitter voltage. The base voltage is generated by the current source that models the generated photo-current. The load resistance is 1 kΩ. b) Generated data generated in COMSOL (dashed lines) compared to a SPICE model (full lines) where the inhibition npn transistor in a) is modeled with Eqs. (2)-(3) with the extracted fit parameters.

## III. MODELING THE NANOWIRE LED

The short lifetime of electron-hole pairs poses a problem also in the design of the nanowire LED. A 5 nm quantum well of InP sandwiched between $Al_{0.3}In_{0.7}P$ is introduced to confine the carriers. This allows for densities high enough for the radiative recombination to overcome the fast trap-assisted recombination at 1.34 ns. As with the npn-phototransistor, the modeling was performed in COMSOL and the same material and simulation parameters were used. The main results from the modeling are summarized in Figs. S4-S5.

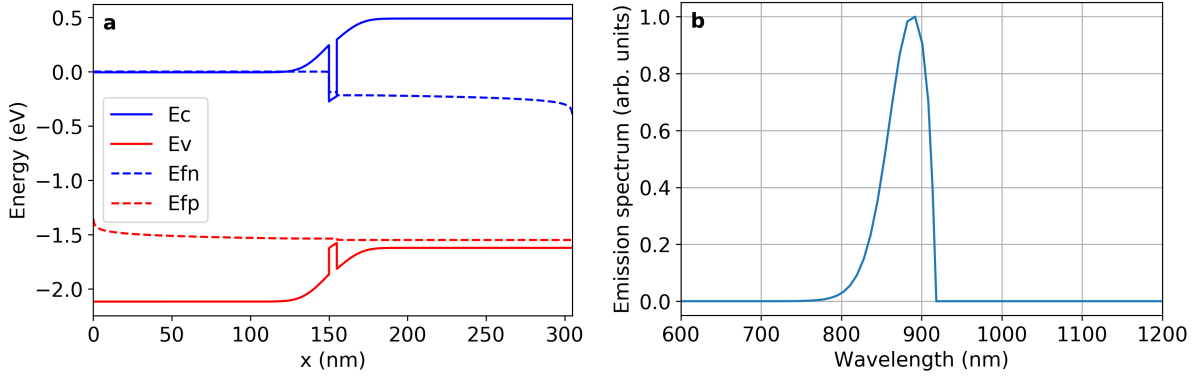

FIG. S4. a) Bandplot of the *pn*-junction with the 5 nm quantum well in the center and b) the corresponding emission spectrum, both at 1.55 V.

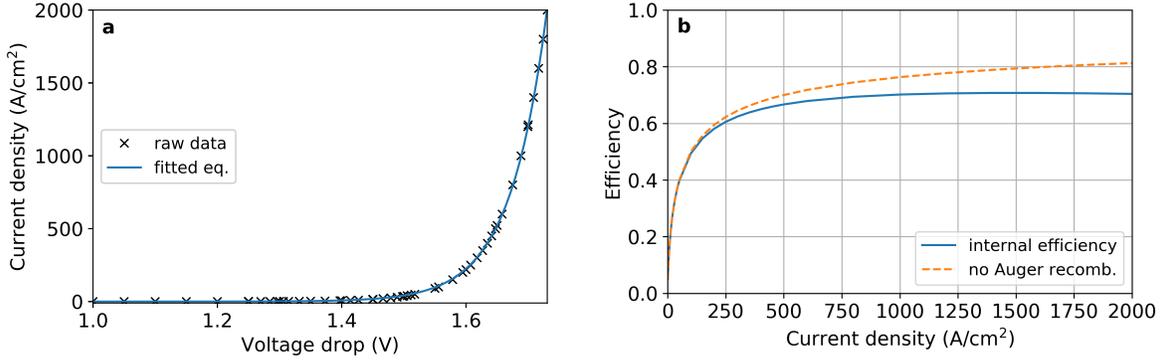

FIG. S5. a) Nanowire LED current density fitted to the Shockley diode equation. b) Internal quantum efficiency modeled with (solid) and without (dashed) Auger recombination.

Our nanowire LED is modeled using the Shockley diode equation

$$J_D(V) = J_S \left( e^{qV/nk_BT} - 1 \right) \quad (6)$$

where $k_BT/q$ is the thermal voltage with $T$ being the temperature of the *pn*-junction. The ideality factor $n$ together with the saturation current density $J_S$ comprise the two fit parameters. Fitting the data yields $J_S = 5.0 \cdot 10^{-10}$ A/cm$^2$ and $n_{\text{eff}} = 2.3$.

## IV. MODELING OF THE SIGMOIDAL COMPONENT AS A CIRCUIT

The full device with two phototransistors and one LED is depicted in Fig. 2 of the main manuscript together with the circuit used to model the device. As explained above, we assume that the circuit parameters for the two npn phototransistors are similar. The load resistance is added to control the load line over the LED and provide a clear saturation limit. To convert the current densities that we acquire from the two fits above we add the information about the wire diameters to the model. In this design 200 nm diameter wires were used for the main body holding the two npn phototransistors, while a thin wire of 50 nm in diameter was used for the branch holding the quantum dot LED. The respective cross-sections are thus $\pi \cdot 10^{-10}$ cm$^2$ and $(\pi/16) \cdot 10^{-10}$ cm$^2$ which is multiplied with the $J_S$ of the photo-transistors and diode, respectively, to find the current equation of each component. The numbers are summarized in Tab. S3 for convenience.

|        | Area                          | $J_S$                          | $I_S$                        |
|--------|-------------------------------|--------------------------------|------------------------------|
| npn PT | $3.14 \cdot 10^{-10}$ cm$^2$  | $5.3 \cdot 10^{-16}$ A/cm$^2$  | $1.7 \cdot 10^{-16}$ nA      |
| QD LED | $1.96 \cdot 10^{-11}$ cm$^2$  | $5.0 \cdot 10^{-10}$ A/cm$^2$  | $9.8 \cdot 10^{-12}$ nA      |

TABLE S3. Cross-section, saturation current density and calculated saturation current for the specific device units for the design discussed here.

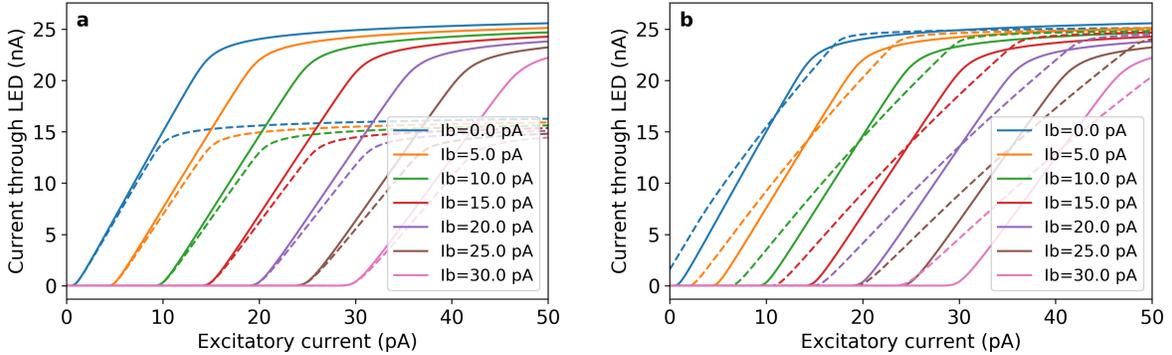

FIG. S6. Different strategies for tuning the activation function. The full circuit in Fig. 2 of the main manuscript is modeled here with the altered parameters a) $V_0 = 3.0$ and $R_{\text{load}} = 50$ MΩ (dashed lines) and b) $V_0 = 5.0$ and $R_{\text{load}} = 110$ MΩ (dashed lines), with the original results using $V_0 = 3.0$ and $R_{\text{load}} = 30$ MΩ (solid lines) as a comparison in both panels.

Using a standard SPICE solver (LTspice), each of the modeled and subsequently parameterized components can be added together in a circuit. Solving for different inhibitory and excitatory currents provides the results given in Fig. 3 of the main manuscript and in Fig. S6 where we give additional examples on how the activation function can be tuned. This func-tion is defined by the onset, saturation level and the slope. In Fig. S6a) the saturation level is tuned by changing the load resistance, and in Fig. S6b) the slope is changed by altering the total bias supply, as well as the applied load resistance.

The slope of the activation function can be tuned in an even wider range if the hardware properties of the device are altered. The forward gain factor the npn phototransistors relates directly to the slope, and can be changed via for example emitter and base doping, base region length and carrier mobility [13], providing a number of options. In Ref. [14], activation functions with a slope value varying with a factor of 2 are used. This can easily be accomplished using any of the approaches listed above. The onset of the activation function could be changed by inserting an additional current source that provides a constant contribution to the exciting current, a concept commonly referred to as a bias unit in neural networks. A simpler way to achieve the same result would be to provide a background excitation and code the signal relative to this background. This is how a constant offset is produced in this work.

## V. ENERGY COST PER OPERATION

In order for the total losses in the network communication not to overcome the transistor gain factor (being the only source of amplification), as discussed in the main text, the emitter must be operated at a relatively high external quantum efficiency. The quantum efficiency is in turn related to the current pushed through the nanowire LED as shown in Fig. S5 b). This effectively creates a minimum power requirement of the device, as the current needed in the LED section needs to be about 100 A/cm$^2$ to perform at an internal efficiency of 50%, which yields $\sim$ 2 nA in the branch. Assuming a few volts across the circuit and additional energy loss due to possible inhibition, a reasonable estimate is a dissipation of 10 nW per device during operation. As argued in the main text, the LED efficiency needs to improved for low current densities to remove this power dissipation bottleneck.

To estimate the operational speed of the neuron device, the device time constant can be calculated following [8], using again the circuit of Fig. 2 of the main text. It is a function of the Shockley emitter resistance $R_e = kT/eI_c$ with $I_c$ as the collector current, the load resistance $R_L$, and the base-emitter and collector-base capacitances $C_e$ and $C_c$, respectively.

It can be represented as

$$\tau = \beta \left[ R_e(C_e + C_c) + R_L C_c \right] \tag{7}$$

where the first term usually dominates. The reason for the RC time constant here being augmented with the forward gain factor $\beta_F$ is the Miller feedback effect [8]. The capacitances (per unit area) are given as [8],

$$C_e = \sqrt{\frac{e^2 \epsilon_e \epsilon_b n_e p_b}{2kT(\epsilon_e n_e + \epsilon_b p_b)(Q_d - v)}}, \tag{8}$$

$$C_c = \sqrt{\frac{e^2 \epsilon_b \epsilon_c p_b n_c}{2kT(\epsilon_b p_b + \epsilon_c n_c)(u - v)}}, \tag{9}$$

where $v$ is the normalized emitter voltage and $u-v$ the collector potential. For this estimate, we will assume $(kT/e)(Q_d - v) = (kT/e)(u - v) = 1$ V which is a common collector potential for our device. Using again a current of 2 nA provides an QD LED internal efficiency of 50% and sets $R_e \approx 1\,\text{M}\Omega$. The parameters for the device, as given above in Tab. S1 for the materials and Tab. S3 for dimensions, yield the capacitances 186 aF and 21 aF for the emitter and collector, respectively. From Eq. (7) a response time of $\sim 6\,\mu\text{s}$ is then found. This number is slightly higher than that reported by [13] which might be due to the use of a large load resistance in our case. A quicker response will require minimizing the emitter capacitance and lowering the load resistance.

To construct a comparable estimate we define here one operation as the entire process of receiving, comparing, processing and outputting new signals. We calculate the energy required per operation by multiplying the on-state power dissipation by the cycle duration. As the power dissipation is fairly constant in the neural node component, the minimal energy required per operation can be found by considering the highest possible operation frequency. Using the same parameters as for the time constant in Eq. (7), the cutoff frequency can be estimated as [13],

$$f_c = \left\{ 2\pi \left[ R_e(C_e + C_c) + R_L C_c \right] \right\}^{-1}, \tag{10}$$

which yields around 100 MHz. Operating at this frequency with 10 nW as on-state power yields an energy dissipation of $10^{-16}$ J per operation, on par with estimates of the human brain energy consumption [15]. At cutoff frequency the gain has however dropped to unity, but the number still serves as an estimate of the current speed limit.

Since real-time operation of the circuit will relate to the sensory input rate, low computational speed can in some cases be desired. The network could operate efficiently within a 50% duty cycle with a pulse duration of $\tau$. This gives the system the possibility to retain information about its previous state in the ring attractor, an important type of short term memory in the ring attractor. The energy consumption per cycle can then be estimated as 10 nW × 6 µs = 60 fJ. Operating the circuit at a lower current density will not affect this estimate much. In order to improve it the time constant needs to be reduced by lower emitter and collector capacitances and lower forward gain values in the npn phototransistors. One way to allow for a lower gain factor would be to decrease the size of the network using instead a stronger control over the radiation patterns. Improvements in the time constant over two orders of magnitude by improving these properties are reasonable, which would reduce the power consumption with the same factor. This brings us down to $\sim 10^{-16}$ J per operation, similar to the high frequency limit above.

To further reduce the power cost of each device, the most important optimization is the LED efficiency at lower currents, as we are far from the low current limit of the phototransistors [16]. This can be done in two ways, either by increasing the radiative recombination rate through Purcell factor engineering as discussed above, or by increasing the non-radiative lifetime of the electron-hole pairs. Using the model of the LED introduced above, we can study the effect of increasing the non-radiative lifetime from the experimental value of 1.34 ns for nanowires to 20 ns for doped bulk InP [17]. This results in a efficiency above 70% for 1 A/cm$^2$ which is two orders of magnitude lower than the current density used for our estimate and would allow us to operate the LED at 0.1 nW. This in turn yields an energy cost per operation of $\sim 10^{-17}$ J (the improvement is limited by the cutoff frequency being affected negatively by a lower current). This represents a considerable potential improvement, especially for low speed or real time computation where the power reduction could be fully utilized.

Additional areas of improvement include miniaturization of the network using nano-antennas, Purcell engineering of the LED using nano-antennas and improvement of the non-radiative lifetimes also in the photo-transistors.

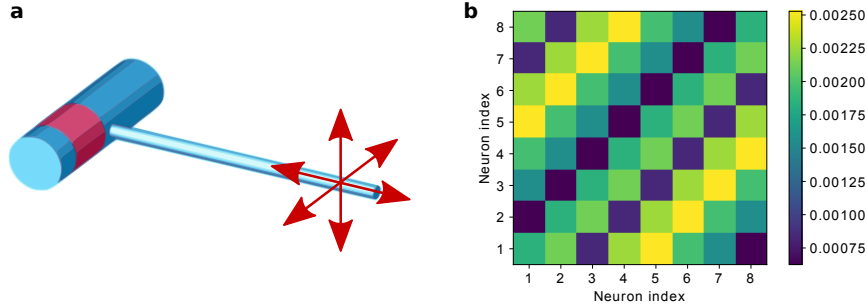

FIG. S7. a) Schematic showing the three orthogonal directions of the dipole source used in the simulations. b) Resulting weight matrix to be compared to Fig. 5c) of the main text.

## VI. ISOTROPIC DIPOLE EMITTER MODEL

In order to asses the importance of polarization selection and the need for an antenna structure, additional modeling of the nanowire LED emission including an isotropic dipole source was carried out. To properly model the isotropic dipole source, three different simulations of the weight matrix were carried out, where the dipole source polarization was varied according to the directions displayed in Fig. S7. After the FDTD simulations [18], the Purcell factor was estimated for each orientation of the source. It was found that the dipole orientations perpendicular to the wire is suppressed to $\sim 30\%$ relative to the orientation along the wire. These results are corroborated by previous theoretical studies [19], where the influence of a semiconductor nanowire on a dipole source has been investigated.

The three simulated weight matrices are weighted according to their respective Purcell factor and added together to the results depicted in Fig. S7b). For this situation with the isotropic emission, the best contrast, i.e. the strongest weights $g_{15}$, were found for a slightly longer wavelength $\lambda = 859$ nm. Plugged in to the computational model of [14], a statistical test was carried out to test the navigational capability, as done in Fig. 6b) in the main text. The results are shown in Fig. S8a).

Although there is some promise in these results, it is not significantly better than the random walk comparable. To improve the results, the ring attractor model can be adjusted to better accommodate the weight matrix of Fig. S7b). Compared to the weight matrix in Fig. 5c of the main text, the sum of all elements is now larger, resulting in a larger signal strength for the mutual inhibition among the ring attractor neural nodes. In the model, there exists a parameter $c$ which balances the excitatory input against the mutual inhibition among these neural nodes (page e5 in [14]). The excitatory and inhibitory inputs are scaled with $c, (1-c)$, respectively. Here we increase this parameter in an attempt to achieve a better signal balance. The results in Fig. S8, are from simulations with a) $c = 0.667$ (original), b) $c = 0.833$ and c) $c = 1.00$, where setting the parameter to unity means canceling the mutual inhibition inside the ring attractor. Comparing the results in Fig. S8a) and b), the network in b) is capable of bringing the agent within 50 steps of the nest for route lengths over 2000 steps for noise levels up to 20%. This is similar to the results in the main text. As stated above, the network in a) did not show successful results. Comparing instead b to c), where mutual inhibition is completely turned off inside the ring attractor, it can be seen that the results are no better than the random walk. It can be concluded that the mutual inhibition is essential for navigation, although for a weight matrix conditioned as the one in Fig. S7b), the internal balancing system needs to be altered in order to put sufficient weight on the excitatory signals. To summarize, we show in this numerical experiment that an isotropic emitter, using only the nanowire itself as an antenna structure, can provide an adequate navigational capability.

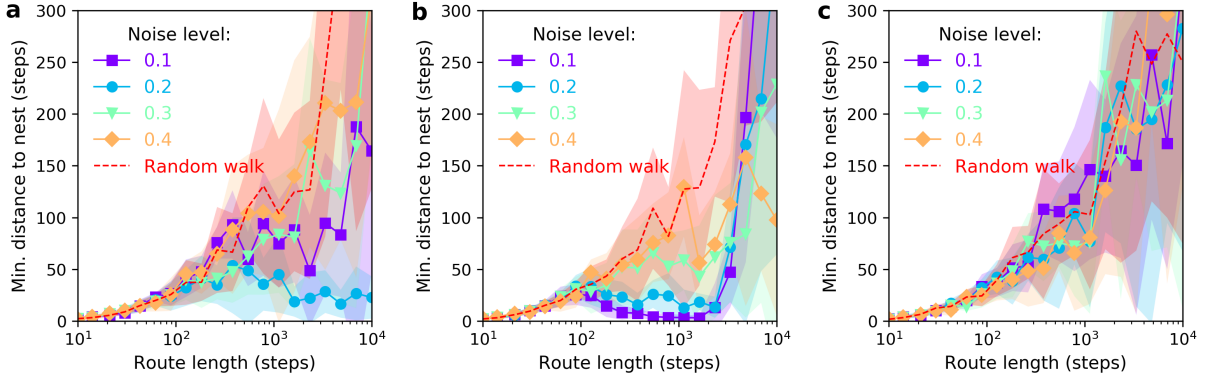

FIG. S8. Statistical tests using the weighting factors: a) $c = 0.667$ (original), b) $c = 0.833$, and c) $c = 1.00$. For each noise level and for the random walk return, 200 trips were carried out distributed on 20 different trip lengths.